# Photonic Floquet Topological Insulators

Mikael C. Rechtsman*[1], Julia M. Zeuner*[2], Yonatan Plotnik*[1], Yaakov Lumer[1], Stefan Nolte[2], Mordechai Segev[1], and Alexander Szameit[2]

[1]*Technion – Israel Institute of Technology, Haifa 32000, Israel*
[2]*Institute of Applied Physics, Abbe Center of Photonics, Friedrich-Schiller-Universität Jena, Max-Wien-Platz 1, 07743 Jena, Germany*

*These authors contributed equally to this work

**The topological insulator is a fundamentally new phase of matter, with the striking property that the conduction of electrons occurs only on its surface, not within the bulk, and that conduction is topologically protected. Topological protection, the total lack of scattering of electron waves by disorder, is perhaps the most fascinating and technologically important aspect of this material: it provides robustness that is otherwise known only for superconductors. However, unlike superconductivity and the quantum Hall effect, which necessitate low temperatures or magnetic fields, the immunity to disorder of topological insulators occurs at room temperature and without any external magnetic field. For this reason, topological protection is predicted to have wide-ranging applications in fault-tolerant quantum computing and spintronics. Recently, a large theoretical effort has been directed towards bringing the concept into the domain of photonics: achieving topological protection of light at optical frequencies. Besides the interesting new physics involved, photonic topological insulators hold the promise for applications in optical isolation and robust photon transport. Here, we theoretically propose and experimentally demonstrate the first photonic topological insulator: a photonic lattice exhibiting topologically protected transport on the lattice edges, without the need for any external field. The system is composed of an array of helical waveguides,**

**evanescently coupled to one another, and arranged in a graphene-like honeycomb lattice. The chirality of the waveguides results in scatter-free, one-way edge states that are topologically protected from scattering.**

The notion of topological protection against scattering known since the experimental discovery of the quantum Hall effect[1–3]. In any ordinary two-dimensional material where time-reversal symmetry is not broken, an electron eigenstate propagating along the edge must have a time-reversed partner associated with it, which propagates in the opposite direction; hence, scattering from imperfections and disorder couples the forward and backward states, giving rise to nonzero resistance. In contrast, in the quantum Hall effect[4], an external magnetic field is applied to a two-dimensional electron gas, breaking time-reversal symmetry, hence the two counter-propagating edge states have different energies, and thus backscattering is prevented. Moreover, the edge state cannot scatter into the bulk of the crystal either, due to the presence of a bulk band gap. These features give rise to perfect edge conductance in the quantum Hall regime, but it requires strong magnetic fields and low temperatures. More recently, it was shown by Kane and Mele[5] that in the case of strong spin-orbit coupling, immunity against scattering (topological protection) can also be achieved without the need for an external field: the materials in this class later became known as topological insulators[6]. In a two-dimensional topological insulator, the mechanism by which topological protection is achieved is Kramers' theorem[5], which disallows backscattering into the counter-propagating state that has opposite electron spin. Another mechanism by which topological protection can be achieved is by applying a time-periodic oscillation in the electron potential, by virtue of an electromagnetic wave passing through the material[7]. It has been conjectured that in such materials, called "Floquet topological insulators," (FTI) spin-orbit coupling is not necessary for achieving topological protection[7,8]. Indeed, the two-dimensional material graphene, which has very weak spin-orbit coupling, has been theorized to exhibit topologically protected edge states under the influence of a circularly polarized electric field[8], but this has not been shown experimentally.

In the domain of optics, Raghu and Handane proposed in 2008 that by using gyromagnetic materials, a photonic analogue of the quantum Hall effect could be realized and topological protection of light could be achieved[9]. Since then, there has been an ongoing effort to realize topological protection in various photonic systems. This is because, apart from their fundamental significance, applications of photonic topological protection include highly robust delay lines[10] as well as efficient and micron-scale on-chip optical isolators[11]. The first observation of topological protection in a photonic system[12] demonstrated the lack of scattering in a microwave, magneto-optic photonic crystal with a strong, external, static magnetic field, similarly to that prescribed by Raghu and Haldane[9]. However, this method of achieving topological protection cannot be carried over to optical frequencies, because magneto-optic effects at optical frequencies are extremely weak. On the other hand, a number of recent works have experimentally demonstrated the existence of photonic edge states that are derived by topological arguments (namely the bulk-edge correspondence principle)[13–15]. However, none of these edge states is topologically protected against scattering and should not be confused with the topologically protected edge states in a topological insulator. This begs the question of how to realize the photonic equivalent of a topological insulator, which should require no external field, be realizable at optical frequencies, and exhibit scatter-free behavior. A number of theoretical efforts have been directed at circumventing the problems arising from the extremely small magnetic response at optical frequencies: (1) employing bi-anisotropic metamaterials with strong magnetic response[16]; (2) using an aperiodic coupled-resonator system that realizes the quantum Hall effect in two uncoupled photonic modes[10]; (3) by dynamically modulating a two-dimensional photonic crystal structure[17] to explicitly break time-reversal symmetry; and (4) by coupling the polarization and propagation degrees of freedom[18]. However, no non-magnetic photonic topological protection has ever been observed experimentally.

Here, we present the first photonic topological insulator. Namely, we experimentally demonstrate the topological protection of transport through photonic edge states in a honeycomb photonic lattice system[14,19–24] (which has been referred to as "photonic graphene[24]"), without any external field whatsoever. The lattice is composed of evanescently coupled helical waveguides, breaking chiral symmetry. In this setting, the propagation of light through the lattice is described by the Schrödinger equation, with the time coordinate replaced by the distance of propagation. In the helical reference frame, propagation is described by an induced fictitious vector potential that is mathematically equivalent to a circularly-polarized, spatially-constant electromagnetic field, acting to break the chiral symmetry. As we show below, the spatial photonic band structure of the structure thus becomes a "Floquet topological insulator," with topologically protected photonic edge states.

The paraxial propagation of light through the photonic lattice is described by the Schrödinger-type equation:

$$i\partial_z \psi(x,y,z) = -\frac{1}{2k_0}\nabla^2 \psi(x,y,z) - \frac{k_0 \Delta n(x,y,z)}{n_0}\psi(x,y,z) \quad (1)$$

where $\psi(x,y,z)$ is electric field envelope function defined by $\mathbf{E}(x,y,z)=\psi(x,y,z)\exp(ik_0 z - i\omega t)\mathbf{x}$; $\mathbf{E}$ is the electric field and $\mathbf{x}$ is a unit vector; $k_0 = 2\pi n_0/\lambda$ is the wavenumber in the ambient medium; $\omega=2\pi c/\lambda$ is the optical frequency; the ambient medium is fused silica with refractive index $n_0=1.45$; and $\Delta n(x,y,z)$ is the refractive index profile of the waveguides (i.e., the deviation from the ambient refractive index in the fused silica). Note that the Schrödinger equation also describes the propagation of a quantum particle wave function that evolves in time (like electrons in a solid, for example). The waveguides act like potential wells, akin to nuclei of atoms in solids. Thus, the diffraction of light through the array of helical waveguides as it propagates in the z-direction is equivalent to the temporal propagation of electrons as they move through a two-dimensional lattice with atoms that are rotating in time.

A microscope image of the input facet of the photonic lattice is shown in Fig. 1(a), and a schematic diagram of the helical waveguides arranged in a honeycomb lattice is shown in Fig. 1(b). The photonic lattice is fabricated using the femtosecond-direct-laser-writing method[25], with each elliptical waveguide having diameters of *11μm* and *3μm* for the horizontal and vertical directions, respectively. The nearest-neighbor spacing between waveguides is *b=15μm*, and the lattice constant is given by *a=26μm*. The increase in index of refraction associated with the waveguides is *7x10⁻⁴*. All waveguide arrays are of length *10cm* in the *z*-direction, which, in the language of the quantum Schrodinger equation, fixes the amount of "time" – *z* – over which the wavepacket evolves. During such evolution in our sample, the wavefunction $\psi$ completes ~20 cycles in phase from *z=0* to *z=10cm*. The helical rotation of the waveguides in our honeycomb lattice is sufficiently slow that a guided mode within a waveguide will be adiabatically drawn along with the waveguide as it rotates. We therefore transform the coordinates into a frame of reference where the waveguides are invariant in the z-direction, namely: *x' = x + Rcos(Ωz); y' = y + Rsin(Ωz); z' = z*, where *R* is the radius of the helix and *Ω=2π/Λ* is the frequency of rotation (*Λ* being the period of rotation). For simplicity, we use the coordinates *(x',y',z')*. In the transformed coordinates, the diffraction of light is given by a modified Schrödinger equation:

$$i\partial_{z'}\psi' = -\frac{1}{2k_0}\left(\nabla' + i\mathbf{A}(z')\right)^2\psi' - \frac{k_0 R^2\Omega^2}{2}\psi' - \frac{k_0\Delta n(x',y')}{n_0}\psi' \quad (2)$$

Where *ψ'= ψ(x',y',z')*, and **A**(z')=*k₀RΩ*(sin(*Ωz'*),-cos(*Ωz'*)) is equivalent to a vector potential associated with a spatially homogeneous electric field of circular polarization. Taken together, the adiabaticity of the guided waveguide modes and the presence of the vector potential allow us to describe the propagation of light through the lattice by coupled mode theory:

$$i\partial_{z'}\psi_n(z') = \sum_{\langle m \rangle} c(\lambda) e^{i\mathbf{A}\cdot\mathbf{r}_{mn}} \psi_m(z') \quad (3)$$

where the summation is only taken over nearest-neighbor waveguides; $\psi_n(z')$ is the amplitude of the mode in the $n^{th}$ waveguide; $c(\lambda)$ is the wavelength-dependent coupling constant between waveguides and $\mathbf{r}_{mn}$ is the spacing between waveguide *m* and waveguide *n*. Here we use the Peierls substitution[26], which yields the phase of the coupling constants between waveguides in the presence of a vector potential. Note that the coupling function c(λ) can be calculated directly by diagonalizing the Schrödinger equation[27]. Since the right-hand side of the equation is z-dependent, there are no static eigenmodes of the system. Rather, the solutions of this equation may be described using Floquet modes, of the form $\psi_n(z')=exp(i\beta z')\phi_n(z')$, where the function $\phi_n(z')$ is $\Lambda$-periodic[7]. This enables the calculation of the spectrum of $\beta$, called the Floquet eigenvalues (also known as the 'quasi-energies' or 'quasi-propagation constants'), as well as the Floquet eigenmodes associated with them. Taken together with Bloch's theorem in the transverse direction, we are able to find the Floquet-Bloch spectrum, or band structure, of the photonic lattice. The band structure for the case of non-helical waveguides *(R=0)* is shown in Fig. 1(c). Plotted as a function of the transverse Bloch wavevector $(k_x,k_y)$, the conical intersections between the first and second bands are the "Dirac points[28]," a key feature of the band structure of graphene. As a result of these, the band structure is analogous to a metallic phase because there is no band gap. When the waveguides are made helical (*R>0*), a band gap in the Floquet spectrum opens, as shown in Fig. 1(d), and the photonic lattice becomes analogous to an insulator - indeed, to a Floquet topological insulator. As we show in the next section, this structure possesses topologically protected edge states.

We calculate the edge band structure, within the tight-binding model (coupled mode theory), by using a unit cell that is periodic in the *x*-direction but finite in the *y*-direction, ending with two "zig-zag" edges (which are therefore treated as infinite in the x-direction). The zig-zag edge is one of three typical edge terminations of the honeycomb lattice, while the other two are the "armchair edge" and the "bearded edge[29]". In our sample (see Fig. 1a), the top and bottom edges are zig-zag edges and the right and left edges are armchair edges. The band structure of the edge states on the zig-zag edge is presented in Fig. 2(a) for the case in which the waveguides are not spinning at all ($R=0$). There are two sets of states residing on the zig-zag edge – one set per edge. Their dispersion curves are entirely flat and completely coincide (i.e., they are degenerate with one another), residing between $k_x = 2\pi/3a$ and $k_x = 4\pi/3a$, occupying one third of $k_x$-space. On the other hand, the Bloch-Floquet band structure for the edge states in a honeycomb lattice of waveguide spinning with $R = 8\mu m$ is shown in Fig. 2(b). In this case the edge states are no longer degenerate with one another, but now have opposite slopes. Specifically, for this case of waveguides spinning in the clockwise direction, the group velocity on the top edge is directed to the right, and on the bottom edge to the left, corresponding to clockwise circulations. At the same time, there are no edge states whatsoever circulating in the counter-clockwise direction, i.e., no edge states on the top edge with group velocity to the left and no edge states on the bottom edge with group velocity to the right. For this reason, these the edge states presented in Fig. 2(b) are exactly the topologically protected edge states of a Floquet topological insulator. The lack of a counter-propagating edge state on a given edge directly implies that any defect or disorder on the edge cannot backscatter light, as there is no backwards-propagating state available into which to scatter, contrary to the case of $R=0$, where there are multiple states into which scattering is possible. This is the very essence of why topological protection occurs in this system. The transverse group velocity of these edge states has a non-trivial dependence on the helical waveguide radius, $R$. For small $R$, the group velocity of the edge states increases, but eventually it reaches a maximum and decreases again. This trend is shown in Fig. 2(c), where we plot the group velocity of the topologically protected edge state at $k_x = \pi/a$ vs. $R$. The maximum calculated group velocity is achieved at $R=10.3\mu m$.

In order to demonstrate the presence of these edge states experimentally, we launch an elliptical beam of light of wavelength *633nm* such that it is incident on the top row of waveguides on the array with radius $R=8\mu m$. The details of the experimental setup are equivalent to those given in a previous work[14]. The position of the input beam is indicated by the ellipse in Fig. 1(a). Experimental data showing the light intensity distribution emerging from the output facet of the lattice is presented in Fig. 3(a) – Fig. 3(d), with the shape and position of the input beam indicated by a yellow ellipse residing on the top row of the lattice. In Fig. 3(a), the output beam emerges at the upper-right corner of the lattice, after having moved along the upper edge. When we move the position of the input beam horizontally slightly to the right, the output beam starts to move down along the vertical right edge of the array, as seen in Fig. 3(b). Clearly, the beam emerging from the lattice remains confined to the lattice edge, not spreading into the bulk of the lattice and without any backscattering. Moving the position of the input beam further rightward results in the output beam moving further down along the side edge, as shown in Fig. 3(c) and Fig. 3(d). Clearly, the input beam has moved along the top edge, encountered the corner, and then continued moving downward along the right edge over the course of propagation. We show this behavior in exact beam-propagation-method (BPM) simulations[30] of the wavepacket as it propagates in the *z*-direction, using the full Schrodinger equation, Eq. (1), as seen in "animation1.gif" of the Supplementary Information. The central observation of these experimental results is that the corner (which can be thought of as a strong defect) does not allow light to scatter. Indeed, no optical intensity is evident along the top edge at the output facet, after having backscattered from the corner. Furthermore, no scattering into the bulk of the array is observed. These observations provide strong evidence of topological protection of the edge state against scattering.

Further evidence of the presence of topological edge states follows from the fact that light stays confined to the side edge of the array as it propagates downwards. This edge is in the armchair edge[29] geometry, which, in a honeycomb lattice composed of straight waveguides ($R=0$) does not allow light confinement at all (i.e., there are no edge states). However, when $R>0$, edge state dispersion calculations (similar to those of Fig. 2(a) and Fig. 2(b)) reveal that a confined edge state emerges following the introduction of non-zero helicity. The fact that light stays confined to the armchair edge, as shown in Fig. 3(a) through Fig. 3(d), proves that indeed making the waveguides helical creates a new edge state on the armchair, which is essential for the topological protection because it prevents transport into the bulk of the lattice.

It is now interesting to experimentally examine the behavior of the topological edge states as the helix radius, $R$, is varied. We find experimentally that the group velocity reaches a maximum and then returns to zero as $R$ is increased, in accordance with Fig. 2(c). To investigate this issue, we fabricate a series of honeycomb lattices of helical waveguides with different R's, and cut them in a large triangle, as displayed in Fig 4. We first examine light propagation in the non-spinning (*R=0*) lattice, which has an equilateral-triangle shape (Fig. 4(a)). We launch a *633nm*-wavelength beam into the waveguide at the upper-left corner of the triangle (circled in yellow). The beam excites two types of spatial eigenstates: (1) bulk states that extend to the corner, and (2) edge states that meet at the corner. As the light propagates in the array, the excited bulk states lead to some degree of spreading into the bulk of the triangle. However, since the edge states are all degenerate (see Fig. 2(a)), they do not cause spreading – even along the edges. Indeed, a superposition of degenerate eigenstates is itself an eigenstate, and so does not change at all during propagation. Figure 4(b) shows the light intensity at the output facet of the lattice, and clearly shows this effect: while some light has diffracted into the bulk of the lattice, the majority remains at the corner waveguide into which the light was launched. This is also shown in a BPM simulation of the diffraction of the beam (where the animation evolves by sweeping through the *z*-coordinate from *z=0cm* to *z=10cm*) in the file 'animation2.gif' of the Supplementary Information.

When the waveguides are made to spin in a clockwise direction around their axis, the edge states are no longer degenerate and have a preferred direction. In fact, the triangular photonic lattice now has a set of edge states that propagate only clockwise on the circumference of the entire triangle. As a result, light at the corner can no longer remain there, and moves along the edge in the clockwise direction. Figs. 4(b) through 4(j) show the light exiting the output facet of the lattice for successively increasing radius, *R*. In the experiments represented by all those figures, the light was launched into the upper-left corner waveguide, marked by the yellow circle in Fig. 4(a). Clearly, increasing $R$ shows a resulting clockwise motion of the wavepacket along the edge. For $R=8\mu m$, the wave packet wraps around the corner of the triangle and moves along the opposite edge, as shown in Fig. 4(f). Importantly, the light is not backscattered even when it hits the acute corner, due to the lack of a counter-propagating edge state. This is a key example of topological protection against scattering. For radius $R=12\mu m$, the wavepacket moves along the edge, but with a slower group velocity (corresponding simulation also shown in 'animation3.gif' of the Supplementary Information – the loss of intensity over the course of propagation is due to bending/radiation losses). This is consistent with the prediction from tight-binding theory that the group velocity of the edge state reaches a maximum at $R=10.3\mu m$ and thereafter decreases with increasing radius. The experimental results suggest that the maximal group velocity is achieved between *6μm* and *10μm*, while the theoretical result of *10.3μm* is well within experimental error, given that this is a prediction purely from coupled-mode (tight-binding) theory. Furthermore, exact BPM simulations confirm this result. By the point at which $R = 16\mu m$, the bending losses are large, leading to a leaking of optical power into scattering modes (which accounts for the large background signal). As shown in Fig. 4(j), the group velocity of the wavepacket approaches zero and therefore the optical power remains at the corner waveguide. These observations clearly demonstrate the presence of a one-way edge state on the boundary of the photonic lattice that behaves according to theory.

In order to show the *z*-dependence of the wavepacket as it propagates along the edge, we turn to a combination of experimental results and simulations where Eq. (1) is numerically evolved according to BPM simulations[30]. In so doing, we examine a waveguide array with a defect on the edge in the form of a "missing" waveguide, as shown in Fig. 5(a). Due to topological protection, the wavepacket should simply propagate around the missing waveguide (i.e., the defect) without backscattering at all. An experimental image of light exiting from the output facet is shown in Fig. 5(b), for a helical array with radius *R=8μm*, at wavelength *λ=633nm*, with propagation distance *z =10cm*. The excited waveguide is at the top left of the triangle, and the edge state propagates in a clockwise sense, avoiding the defect and propagating through it, eventually hitting the next corner of the triangle. It is instructive to actually follow the evolution of the wavepacket as it propagates, hence we show simulation results of the intensity of the wavepacket at intervals of *2cm* during the propagation through the lattice. In Fig. 5(c) through Fig. 5(h), we show simulation results for the optical intensity at *z=0cm, z=2cm, z=4cm, z=6cm, z=8cm, and z=10cm*. The wavepacket is launched into the waveguide at the top-right corner (circled with yellow in Fig. 5), and it clearly propagates around the defect and continues forward without backscattering. The presence of weak optical intensity in the bulk of the lattice arises from overlap of the initial launch beam with the bulk modes of the system. Note that the wavepacket in the simulations has progressed slightly farther along the edge than that in the experimental results. This is a result of a small degree of uncertainty in the value of the coupling constant, $c(\lambda)$. Taken together, these data show the longitudinal progression of topologically protected modes as they travel along the edge of the lattice.

Photonic FTIs enable topological protection of light waves in the optical regime, and without an external field. This potentially leads to an entirely new platform upon which topological protection, a general wave effect clearly not confined to just solid-state systems, can be understood and probed. For example, our photonic lattices have the same geometry as photonic crystal fibers, and thus these systems are likely to exhibit robust topologically protected states. Many interesting open questions are prompted: what will be the behavior of entangled photons in a topologically protected system? Upon the introduction of nonlinearity, what will be the effect of interactions on the non-scattering behavior? Can photonic Majorana fermions be realized, for applications in robust quantum computing? The realization of a photonic FTI in our relatively simple classical system will enable these questions, as well as many others, to be addressed.

Acknowledgements:

M.C.R. is grateful to the Azrieli foundation for the Azrieli fellowship while at the Technion. A.S. gratefully acknowledges the support of the German Ministry of Education and Research (Center for Innovation Competence program, grant 03Z1HN31). M.S. gratefully acknowledges the support of the Israel Science Foundation, the USA-Israel Binational Science Foundation, and the Advanced Grant by the European Research Council. The authors thank D. Podolsky, S. Raghu and T. Pereg-Barnea for helpful discussions.

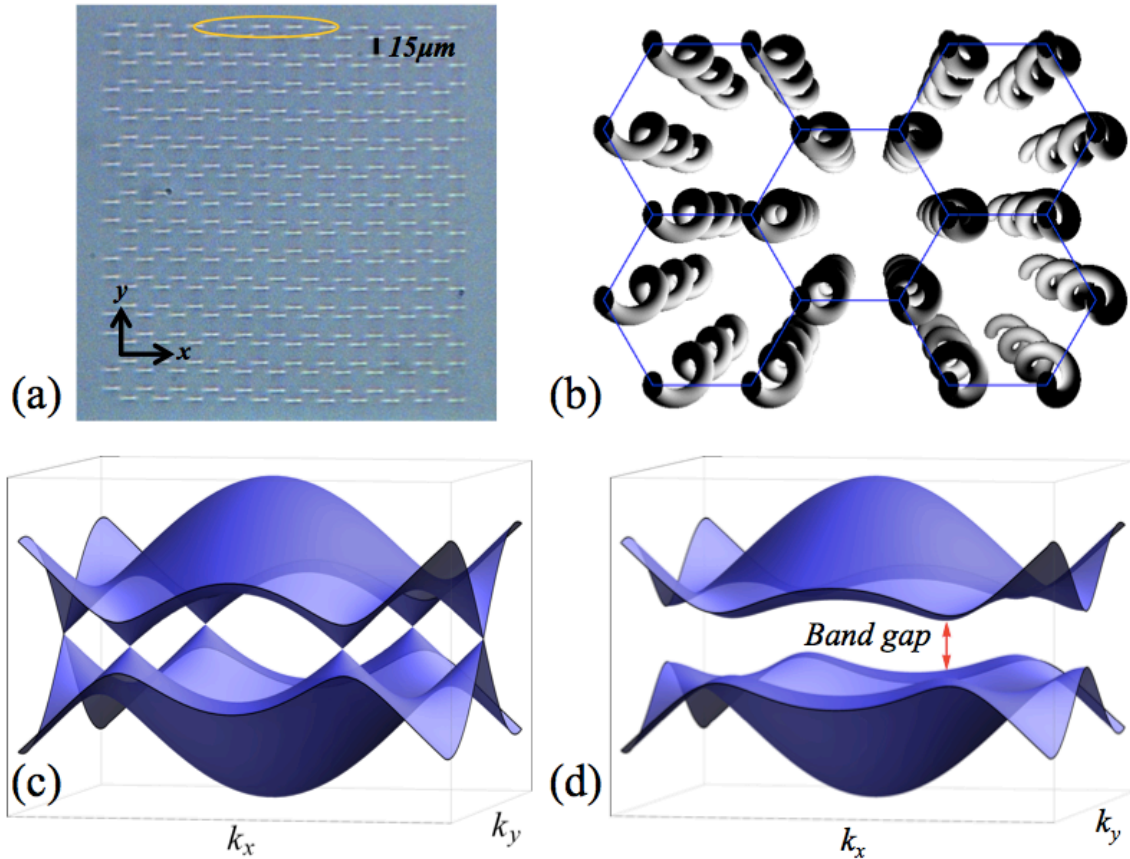

Fig. 1: **Geometry and band structure of honeycomb photonic Floquet topological insulator lattice.** (a) Input facet of photonic lattice, honeycomb geometry with "zig-zag" edge terminations on the top and bottom, and "armchair" terminations on the left and right sides. (b) Schematic diagram of the helical waveguides. The waveguides are helical with their rotation axis in the z-direction, with radius R and pitch Z. (c) Spatial band structure ($\beta$ vs. $(k_x,k_y)$) for the case of non-helical waveguides comprising a honeycomb lattice (R=0). Note the band crossings at the Dirac point. (d) Spatial bulk band structure for the photonic topological insulator: helical waveguides with R=8μm arranged in a honeycomb lattice. Note the band gap opening up at the Dirac points (labeled with the red, double-sided arrow), which corresponds to the band gap in a Floquet topological insulator.

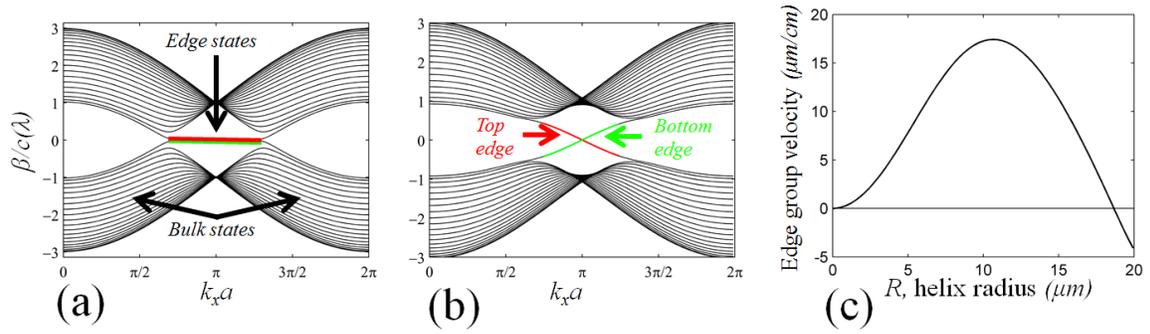

Fig. 2: Dispersion curves of the edge states, highlighting the unique curves of the topologically protected edge states for the helical waveguides in the honeycomb lattice.
(a) Spatial band structure of the edge states on the top and bottom of the array when the waveguides are straight (R=0). The dispersion of both top and bottom edge states (red and green curves) is flat, therefore they have zero group velocity. The bands of the bulk honeycomb lattice is drawn in black. (b) Dispersion curves of the edge states in the Floquet topological insulator for helical waveguides with R = 8μm arranged in the honeycomb lattice: the band gap opens and the edge states acquire nonzero group velocity. These edge states reside strictly within the bulk band gap of the bulk lattice (drawn in black). (c) Group velocity (slope of green and red curves) vs. radius of rotation, R, of the spinning waveguides comprising the honeycomb lattice. The maximum occurs at *R=10.3μm*.

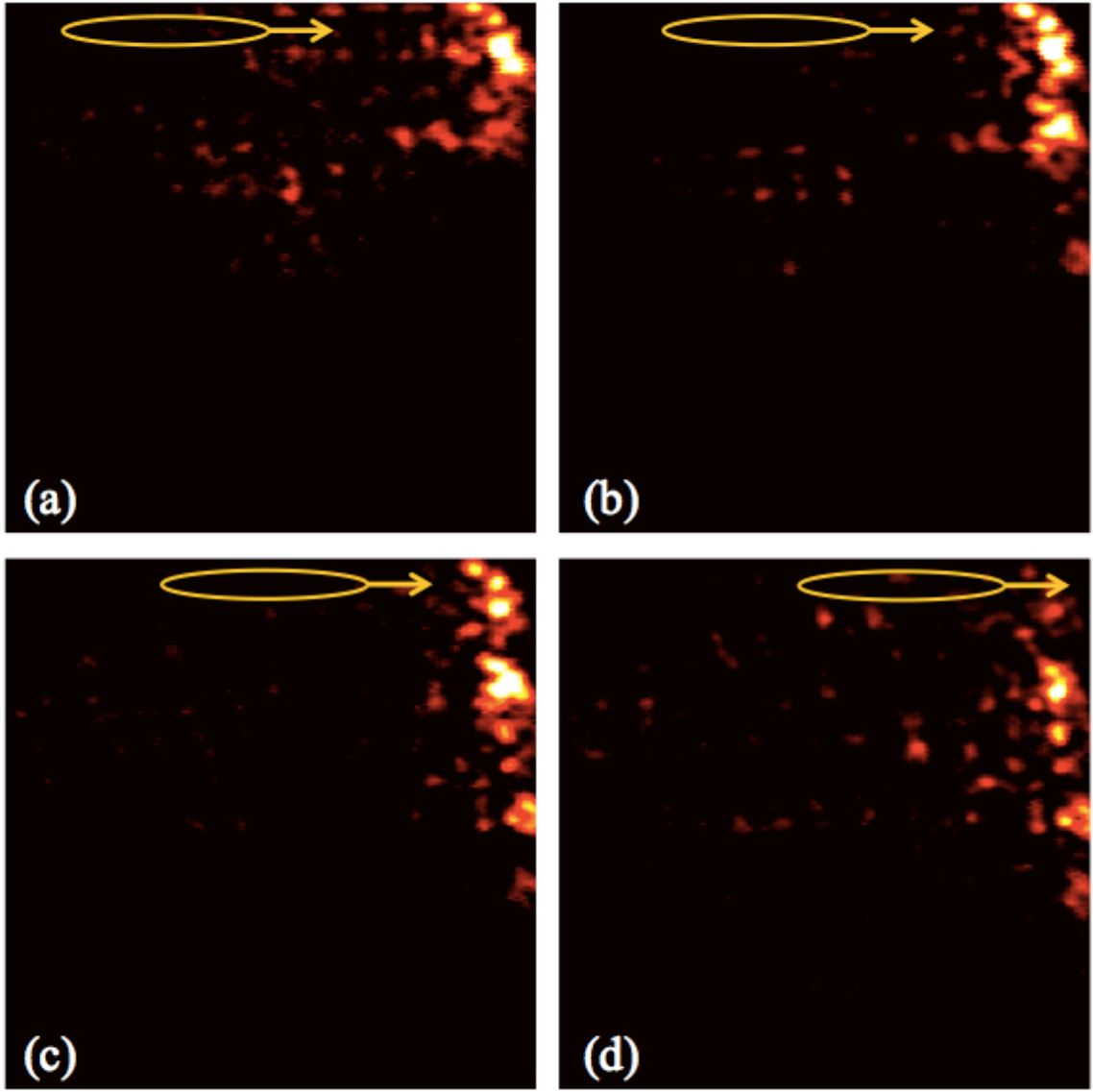

Fig. 3: Light emerging from the output facet of the waveguide array as the input beam is moved rightward, from (a) to (d), along the top edge of the waveguide array. The beam propagates along the top edge of the array (which is in the zig-zag configuration), hits the corner, and clearly moves down the vertical edge (which is in the armchair configuration). Note that the wavepacket shows no evidence of backscattering or bulk scattering due to its impact with the corner of the lattice. This scattering of the edge state is prevented by topological protection.

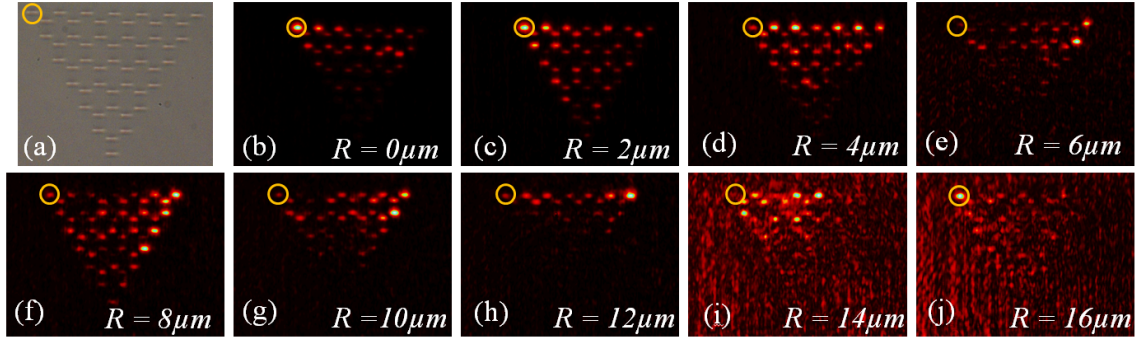

Fig. 4: **Experiments highlighting light circulation strictly along the edges of triangularly-shaped lattice of helical waveguides arranged in a honeycomb geometry.** (a) Microscope image of the honeycomb lattice in the triangular configuration. (b)-(i=j) Light emerging from the output facet of the photonic lattice (after *10cm* of propagation) for increasing helical radius, *R*, at wavelength *633nm*. The light is initially launched into the waveguide at the upper-left corner, surrounded by a yellow circle. At *R=0*, the initial beam excites a confined defect mode at the corner. As the radius is increased, light is moving along the edge by virtue of a topological edge mode. It reaches its maximum displacement near *R = 8µm*. The light wraps around the corner of the triangle and is not backscattered at all: this is a clear example of topological protection against scattering. The light slows down and finally stops near R = 16µm. The large degree of background noise in (i) and (j) is due to high bending losses of the waveguides as a result of coupling to free-space scattering modes.

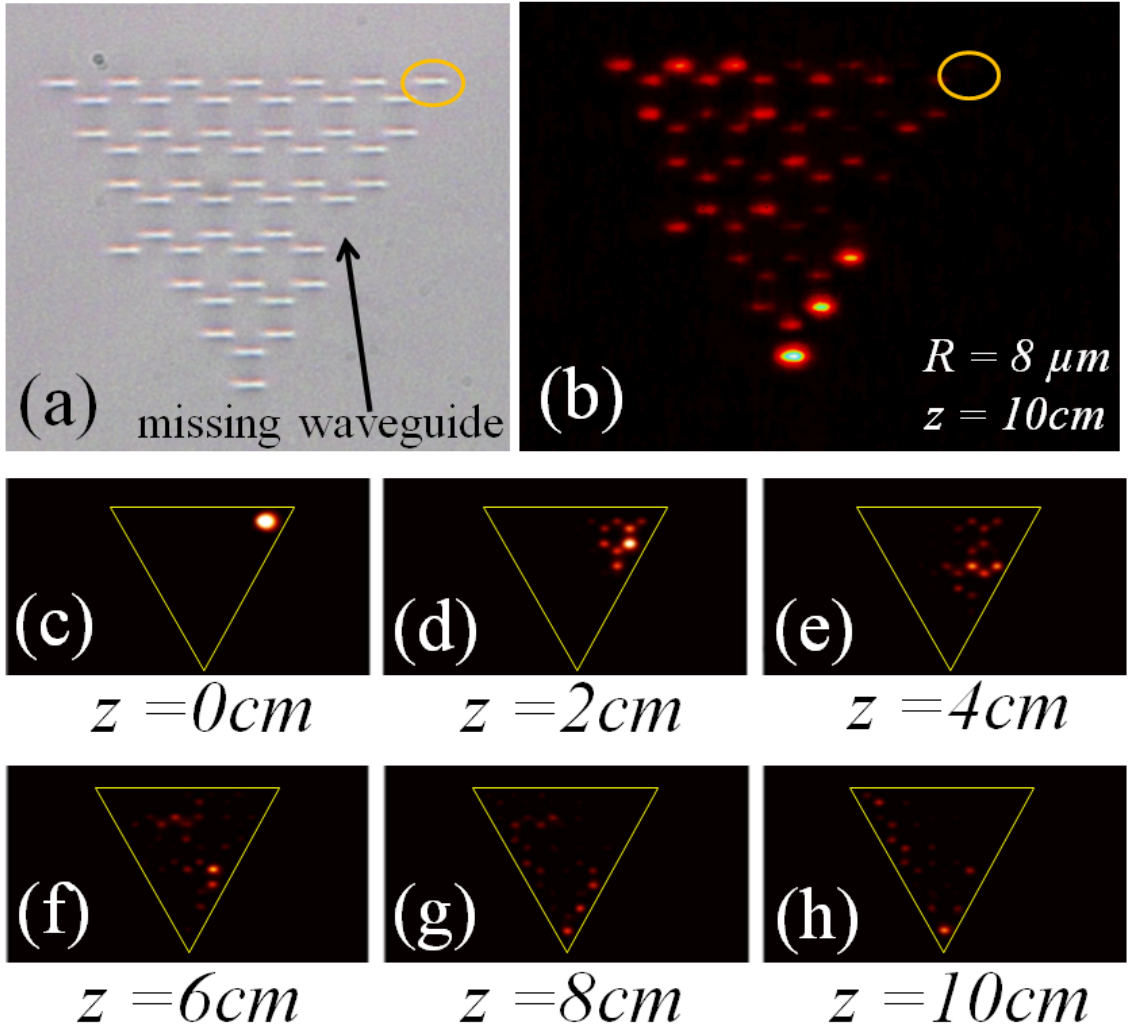

Fig. 5:
**Experiments and simulations showing topological protection in the presence of a defect.** The lattice is triangularly-shaped, and the waveguides are helically spinning with *R=8μm*. (a) Microscope image of photonic lattice with missing waveguide (acting as a defect) on the right-most zig-zag edge. A light beam of *λ=633nm* is launched into the single waveguide at the upper-right corner (surrounded by a yellow circle). (b) Experimental image of light emerging from the output facet after *z=10cm* of propagation, showing no backscattering despite the presence of the defect (a signature of topological protection). (c)-(h) Numerical simulation of light propagation through the lattice at various propagation distances (*z=0cm, 2cm, 4cm, 6cm, 8cm and 10cm,* respectively). After minimal bulk scattering, the light propagates along the edge, encounters the defect, propagates around it, and continues past it without scattering, in agreement with (b).